\documentclass[preprintnumbers,superscriptaddress,showkeys,byrevtex]{revtex4}
\usepackage{amsmath,amsfonts,amssymb,amscd,amsxtra,amsthm}
\usepackage{graphicx}
\usepackage{bm}
\usepackage{subcaption}
\usepackage{xcolor}
\usepackage{orcidlink}
\begin{document}
\preprint{PKNU-NuHaTh-2023}
\title{Study on the $\phi$-meson photoproduction off the proton target \\with the pentaquark-like $K^*\Sigma$ bound state $P_s$}
\author{Sang-in Shim\,\orcidlink{0000-0003-3161-1018}}
\email[E-mail: ]{shimsang@rcnp.osaka-u.ac.jp}
\affiliation{Center for Extreme Nuclear Matters (CENuM), Korea University, Seoul 02841, Korea}
\affiliation{Department of Physics, Pukyong National University (PKNU), Busan 48513, Korea}
\author{Yongsun Kim\,\orcidlink{0000-0002-9025-0489}}
\email[E-mail: ]{yongsun@sejong.ac.kr}
\affiliation{Department of Physics, Sejong University, Seoul 05006, Korea}
\affiliation{Center for Extreme Nuclear Matters (CENuM), Korea University, Seoul 02841, Korea}
\author{Seung-il Nam\,\orcidlink{0000-0001-9603-9775}}
\email[E-mail: ]{sinam@pknu.ac.kr, gariwulf@gmail.com}
\affiliation{Department of Physics, Pukyong National University (PKNU), Busan 48513, Korea}
\affiliation{Center for Extreme Nuclear Matters (CENuM), Korea University, Seoul 02841, Korea}
\affiliation{Asia Pacific Center for Theoretical Physics (APCTP), Pohang 37673, Korea}
\date{\today}
\begin{abstract}
We utilize the effective Lagrangian method within the tree-level Born approximation to explore $\phi$-meson photoproduction, i.e., $\gamma p \to \phi p$. 
Our analysis encompasses contributions from various sources, including the Pomeron, $f_1$-Regge, pseudoscalar particles ($\pi$, $\eta$), scalar particles ($a_0$, $f_0$), protons, and three-nucleon resonance states. 
In addition, we consider a possible pentaquark-like $K^* \Sigma$-bound state $P_s$. 
The findings indicate that, apart from the region near the threshold, contributions other than the Pomeron generally have a limited impact on the total cross section. 
However, at specific angles, alternative contributions become crucial, particularly at smaller values of ${\rm cos}\,\theta$.
The incorporation of $P_s$ and other nucleon resonances proves essential to elucidate the bump observed near $W \sim 2.15$ GeV at very forward angles and behaviors within the range of $W=(2.0-2.3)$ GeV. 
Furthermore, in the region with $W\ge 2.5$ GeV, where nucleon resonances become negligible, contributions from the t-channel mesons become pivotal. 
Our calculations for spin density matrix components, examined in various frames, exhibit improvement when considering all contributions. 
This comprehensive approach successfully reproduces the observed bump by including $P_s$.
We also briefly estimate the $P_s$ production via $\phi$-meson photoproduction in the future Electron-Ion Collider (EIC),
resulting in the luminosity of 10 fb$^{-1}$ per month.
\end{abstract}
\maketitle
\section{Introduction}
Research on $\phi$-meson photoproduction has been extensive, encompassing both theoretical investigations and experimental inquiries. 
This phenomenon serves as a valuable tool for gaining insights into the properties of the strong interaction and structure of hadrons. 
Early theoretical explorations of $\phi$-meson photoproduction focused on the Pomeron exchange mechanism, a non-resonant process primarily driven by the exchange of colorless gluons. 
While this mechanism is believed to account for the overall energy dependence of the cross section, it falls short of explaining the observed bump structure near the threshold. 
In 2005, the LEPS Collaboration made a significant observation, reporting a distinctive bump structure in the total cross section for $\phi$-meson photoproduction near the threshold~\cite{LEPS:2005hax}. 

Despite consistent observations of this structure in subsequent experiments~\cite{CLAS:2013jlg,Dey:2014tfa} and theoretical studies~\cite{Ozaki:2009mj,Kiswandhi:2010ub,Kiswandhi:2011cq,Ryu:2012tw}, its origin remains a perplexing question. 
A clue to this puzzle may be found in exotic hadrons.  
After the recent observation of the pentaquarks $P_c(4312)$, $P_c(4440)$, $P_c(4450)$ at the Large Hadron Collider (LHC)~\cite{LHCb:2019kea}, additional theoretical and experimental investigations are underway. 
Among them, there are investigations into $J/\psi$-photoproduction involving the consideration of $P_c(4450)$ with a mass similar to the threshold of $D^*\Sigma_c$~\cite{Wang:2015jsa,HillerBlin:2016odx,Wang:2019krd,GlueX:2019mkq}. 
Substituting charm contents with strange ones in this reaction leads to $\phi$-photoproduction, considering pentaquarks with a mass similar to the $K^*\Sigma$ threshold. 
Theoretical studies on $\phi$-photoproduction, including nucleon resonances with masses similar to the $K^*\Sigma$ threshold and having the same spin-parity as their bound state, have been conducted~\cite{Kiswandhi:2010ub,Kiswandhi:2011cq}. 
However, in these works, the couplings between the resonance states, photon, and $\phi$-meson, as well as the widths of the resonances, are treated as free parameters. 
The $K^*\Sigma$ molecular bound-state was already suggested using the coupled-channel method including the mass, coupling with proton and $\phi$-meson, and the width~\cite{Khemchandani:2011et}. 
Based on this, an investigation on hidden-strangeness production, such as $K^+p \to K^+ \phi p$, has also been conducted~\cite{Nam:2021ayk}.

In the present work, we investigate the role of the $K^* \Sigma$-bound state (we denote it as $P_s$) in the $\phi$-meson photoproduction. 
$P_s$ is a poorly known state with $S=0$ and $J^P=3/2^-$ suggested by the unitarized coupled-channel meson-baryon interaction model~\cite{Khemchandani:2011et}. 
If proven to exist this structure would play a role in various hadronic reactions, including the process of $\phi$-meson photoproduction. 
To explain the total (Refs.~\cite{Ballam:1972eq,Barber:1981fj,Egloff:1979mg,Busenitz:1989gq}) and differential (Ref.~\cite{Dey:2014tfa}) cross section of $\phi$-meson photoproduction, we employ a comprehensive theoretical framework 
that considers contributions
from various components, including the Pomeron exchange, t-channel meson exchanges, and the direct $\phi$-radiation via the proton and nucleon resonances including $P_s$. 
We determine the relevant parameters by fitting the model to the experimental data. 
In addition to the total and differential cross sections, our analysis also discusses the spin-density matrix elements (SDMEs).

The numerical results indicate that contributions aside from the Pomeron are typically insignificant regarding the total cross section, except especially near the threshold. 
As energy levels increase, contributions other than the Pomeron and $f_1$ Regge are considered insignificant. 
However, when examining production reactions at specific angles rather than the total cross section, relying solely on the Pomeron contribution proves insufficient. 
Hence, it becomes essential to analyze additional contributions. 
We calculate $d\sigma /d {\rm cos}\,\theta$ to evaluate various angles and center-of-mass (cm) energies, comparing them with CLAS data. 
It is clear that the Pomeron contribution adequately explains the experimental data in the forward-scattering region. 
However, as ${\rm cos}\,\theta$ increases, the deviation from the data increases, indicating the need for contributions other than the Pomeron, especially at larger ${\rm cos}\,\theta$ values.
Importantly, including $P_s$ and other nucleon resonances is crucial for accurately describing the bump observed near $W \sim 2.15$ GeV at very forward angles, as well as for capturing behaviors within the range of $(2.0-2.3)$ GeV. 
Here, $W$ denotes the total energy in the center-of-mass (cm) frame. 
Additionally, in the region for $W \ge 2.5$ GeV, where nucleon resonances are scarce, contributions from $t$-channel mesons, separate from the Pomeron, become necessary.

We also compute the SDME $\rho^0_{00}$ at ${\rm cos}\,\theta = 0.7$ in the Gottfried-Jackson, Adair, and Helicity frames, and compare them with CLAS data. 
When considering only the Pomeron, the results consistently underestimate the data across all frames. 
However, by including all contributions, this disparity is significantly reduced, resulting in a better alignment between the data and the model. 
An important achievement of this study is the successful replication of the observed bump at $W \sim 2.15$ GeV by incorporating $P_s$. 
Our findings suggest that the upcoming Electron-Ion Collider (EIC) experiment will facilitate precise measurements of the $P_s$ resonance. 
We estimate the $P_s$ production via $\phi$-meson photoproduction in the EIC.

The remainder of this paper is organized as follows. 
In Section II, we outline the framework and provide numerical inputs for the reaction. 
Section III is devoted to comparing our numerical results with existing data and discussing them, while the final section offers a summary and conclusion.

\section{Theoretical Framework}
\label{sec:II}
In this Section, we would like to describe the theoretical framework briefly for the $\phi$-photoproduction:
\begin{equation}
\gamma(k_1) + p(p_1) \rightarrow \phi(k_2) + p(p_2).
\label{eq:reac}
\end{equation}
In the cm frame, the momenta of the particles can be defined as follows:
\begin{eqnarray}
\label{eq:}
k_1 &=& (k, 0, 0, k),\cr
p_1 &=& (\sqrt{k^2 + M_N^2}, 0, 0, -k), \cr
k_2 &=& (\sqrt{k'^2 + M_{\phi}^2}, k'{\rm sin}\,\theta, 0, k'{\rm cos}\,\theta),\cr
p_2 &=& (\sqrt{k'^2 + M_N^2}, -k'{\rm sin}\,\theta, 0, -k'{\rm cos}\,\theta).
\end{eqnarray}
Here, the $z$-axis is set to be parallel to the direction of the photon and the $y$-axis is normal to the reaction plane. 
The masses of the proton and the $\phi$-meson are denoted by $M_N$ and $M_{\phi}$, respectively. 
The angle between the directions of the initial photon and the final $\phi$-meson is given by $\theta$. 
The magnitudes of the initial and final three momenta $k$ and $k'$ are defined as
\begin{eqnarray}
\label{eq:}
    k &=& (W^2 - M_N^2)/(2W),\cr
    k' &=& \sqrt{(M_{\phi}^2)^2+(M_N^2)^2+(W^2)^2 - 2(M_{\phi}^2 M_N^2 + M_{\phi}^2 W^2 + M_N^2 W^2 )}/(2W),
\end{eqnarray}
where $W$ is the total energy of the system in the cm frame.

The invariant amplitude can be written as 
\begin{equation}
\mathcal{M} = \varepsilon_{\nu}^{*}(k_2,\lambda')
 \bar{u} (p_2,s')\mathcal{M}^{\mu\nu}(k_1,p_1,k_2,p_2) u (p_1,s)
\epsilon_{\mu}(k_1,\lambda)
\label{eq:invAmp}
\end{equation}
where $\epsilon_{\mu}(k_1,\lambda)$ and $\varepsilon_{\nu}^{*}(k_2,\lambda')$ are the polarization vector of the photon and $\phi$-meson, respectively. $u(p_1,s)$ and $u(p_2,s')$ are the spinors of the initial- and final-state protons, respectively.

The invariant amplitude of Eq.~\eqref{eq:invAmp} consists of the contributions from $t$-channel Pomeron exchange, $t$-channel pseudo-scalar ($\pi$, $\eta$), scalar ($a_0(980)$, $f_0(980)$), axial-vector ($f_1(1285)$) meson exchange, $s$- and $u$-channel nucleon $N$, and nucleon resonance $N^*$ exchange as follows:
\begin{equation}
    \mathcal{M} = \mathcal{M}_{\mathbb{P}}+\mathcal{M}_{\rm PS}+\mathcal{M}_{\rm S}+\mathcal{M}_{\rm AV}+\mathcal{M}_{N}+\mathcal{M}_{N^*}
    \label{eq:allAmp}
\end{equation}

\begin{figure}[t]
\includegraphics[width=15cm]{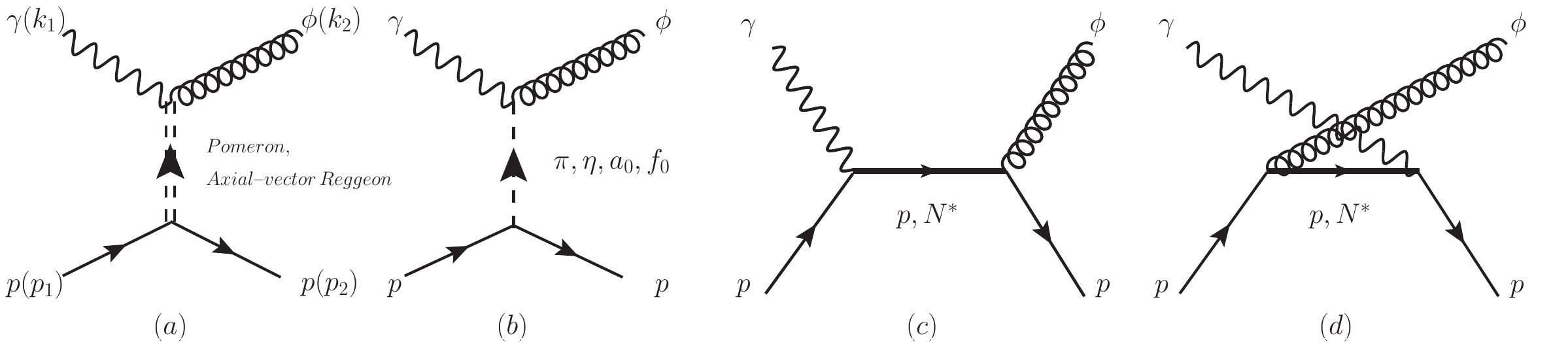}
\caption{Relevant contributions from (a) $t$-channel Pomeron exchange, (b) $t$-channel pseudo-scalar ($\pi$, $\eta$), scalar ($a_0(980)$, $f_0(980)$), axial-vector ($f_1(1285)$) meson exchange, (c, d) $s$- and $u$-channel nucleon $N$, and nucleon resonance $N^*$ exchange}
\label{fig:01}
\end{figure}

In Fig.~$\ref{fig:01}$, Feynman diagrams for each contribution are drawn. For the nucleon resonance $N^*$, we consider only three resonace states, $N^*(2000,5/2^+)$, $N^*(2300,1/2^+)$, and the pentaquark-like $K^* \Sigma$-bound state $P_s$, which is assigned by $J^P=3/2^-$ from the coupled-channel $V$-$B$ interaction model, in which $V$ and $B$ denote the vector meson and baryon, respectively~\cite{Khemchandani:2011et}.

\subsection{Pomeron exchange}
In what follows, we introduce the effective Lagrangians of the various contributions for the $\phi$ photoproduction.  First, for the calculation of Pomeron-exchange contribution, we consider the $C=+1$ vector-like Pomeron ($\mathbb{P}$), following Refs.~\cite{Titov:1999eu,Titov:2003bk,Kim:2019kef,Kim:2020wrd}. Its amplitude is depicted in Fig.~\ref{fig:01}(a). Taking into account the VMD model, we can construct the following effective Lagrangians:
\begin{eqnarray}
    \mathcal{L}_{\gamma\mathbb{P} \phi } 
    &=&i g_{\gamma \mathbb{P}\phi }
    F_{\phi}(t) \left[ 
        ( \phi_{\nu} \partial^{\mu} \mathbb{P}^{\nu} 
         -\mathbb{P}_{\nu} \partial^{\mu} \phi^{\nu} )A_{\mu}
        -(A_{\nu}\partial^{\mu}\mathbb{P}^{\nu} 
         -\mathbb{P}_{\nu}\partial^{\mu}A^{\nu} 
        )\phi_{\mu}
        -(\phi_{\nu}\partial^{\mu} A^{\nu} - A_{\nu}\partial^{\mu}\phi^{\nu}
        )\mathbb{P}_{\mu}
    \right],\cr
    \mathcal{L}_{\mathbb{P} NN} 
   & =&g_{\mathbb{P} NN } F_{N}(t) \bar{N} \rlap{/}{\mathbb{P}} N,
\label{eq:LagPo}
\end{eqnarray}
where $N$, $\mathbb{P}^{\mu}$, $A^{\mu}$, and $\phi^{\mu}$. The invariant amplitude derived from Eq.~\eqref{eq:LagPo} can be written as 
\begin{eqnarray}
    \mathcal{M}_{\mathbb{P}}&=&
    \mathcal{\varepsilon^{*}_{\nu}(\lambda')}\bar{u}_{N}(\lambda_{f})
    \mathcal{M}^{\mu\nu}_{\mathbb{P}}
    u_{N}(\lambda_i)\epsilon_{\mu}(\lambda),
\cr
    \mathcal{M}_{\mathbb{P}}^{\mu\nu} &=&
    -M_{\mathbb{P}}(s,t)
    \left[
        \left(
            g^{\mu\nu} - \frac{k_2^{\mu} k_2^{\nu} }{k_2^2 }
        \right)\rlap{/}{k}_1
        -
        \left(
            k_1^{\nu} - \frac{k_2^{\nu} k_1\cdot k_2}{k_2^2}
        \right)\gamma^{\mu}
        -
        \left(
            \gamma^{\nu} - \frac{\rlap{/}{k}_2 k_2^{\nu} }{k_2^2}
        \right)
        \left(
            k_2^{\mu} - \frac{\bar{p}^{\mu} k_1\cdot k_2}{\bar{p}\cdot k_1}
        \right)
    \right].
    \label{eq:MPmunu}
\end{eqnarray}
Here, $\bar{p}=(p_1 + p_2)/2$. The last term $\propto\bar{p}$ in the square bracket is added to restore the gauge invariance~\cite{Titov:2003bk}. The scalar factor $M_{\mathbb{P} }(s,t)$ in Eq.~\eqref{eq:MPmunu} is given by
\begin{equation}
    M_{\mathbb{P} } (s,t)
    = C_{\mathbb{P} } F_{\phi}(t) F_{N}(t)
    \frac{1}{s} \left( \frac{s}{s_{\mathbb{P} } } \right)^{\alpha_{\mathbb{P} }(t) }
    \exp\left[
        -\frac{i\pi}{2}\alpha_{\mathbb{P} }(t)
    \right]. 
\end{equation}
Here, we define the strength factor $C_\mathbb{P} = g_{\gamma\mathbb{P}\phi } g_{\mathbb{P} NN } = 3.6$~\cite{Titov:2003bk, Kim:2020wrd} for convenience and the energy-scale factor $s_{\mathbb{P} } = ( M_{N} + M_{\phi} )^2$. The Pomeron trajectory is known to be $\alpha_{\mathbb{P} }(t) = 1.08 + 0.25 t$. The form factor for the $\gamma \mathbb{P} \phi$ vertex $F_{\phi}(t)$~\cite{Jaroszkiewicz:1974ep,Donnachie:1983hf} and the form factor for the nucleon $F_{N}(t)$~\cite{Laget:1994ba,Donnachie:1988nj} take the forms
\begin{eqnarray}
    F_{\phi}(t)&=&
    \frac{2 \mu_0^2 \Lambda_{\phi}^2}{
         (\Lambda_{\phi}^2 - t )( 2\mu_{0}^2 + \Lambda_{\phi}^2 - t ) },
  \cr
    F_{N}(t)&=& 
    \frac{ 4 M_{N}^2 - 2.8 t }{ 
         ( 4 M_{N}^2 - t )( 1 - t/t_0 )^2 }.
\end{eqnarray}
where the scale parameters and the cutoff are given by $\mu_0^2=1.1\, {\rm GeV}^2$, $t_0=0.71\,{\rm GeV}^2$, and $\Lambda_{\phi}=4 M_{\phi}^2$.

\subsection{Meson exchange}
For the meson-exchange contributions, shown in Fig.~\ref{fig:01}(b), we consider the Reggeized axial-vector meson ($f_1$-Regge) related to $f_1(1285)$ which has the quantum number $I^G(J^{PC})=0^+(1^{++})$~\cite{Donnachie:2002en, Kochelev:1999zf}, pseudoscalar mesons, and scalar mesons. The effective Lagrangians for the mesonic vertices are written as follows:
\begin{eqnarray}
    \mathcal{L}_{f_1 \gamma \phi}
    &=&g_{f_1 \gamma \phi} \epsilon^{\mu\nu\alpha\beta}
    \partial_{\mu}A_{\nu}\partial^2\phi_{\alpha}f_{1\beta},\cr
    \mathcal{L}_{\Phi \gamma \phi}
    &=& \frac{e g_{\Phi \gamma \phi}}{M_{\phi} }
    \epsilon^{\mu\nu\alpha\beta}
    \partial_{\mu}A_{\nu}\partial_{\alpha}\phi_{\beta}\Phi,\cr
    \mathcal{L}_{S \gamma \phi}
    &=& \frac{e g_{S \gamma \phi}}{M_{\phi} }
    F^{\mu\nu}\phi_{\mu\nu}S,
    \label{eq:LagMVV}
\end{eqnarray}
and, for the nucleon vertices,
\begin{eqnarray}
    \mathcal{L}_{f_1 NN}
    &=&-g_{f_1 NN} \bar{N}
    \left[
        \gamma_{\mu} 
        - i\frac{\kappa_{f_1 NN}}{2M_N}
        \gamma_{\nu}\gamma_{\mu}\partial^{\nu}
    \right]
    f^{\mu}_1 \gamma_5 N, \cr
    \mathcal{L}_{\Phi NN}
    &=& i g_{\Phi NN} \bar{N} \Phi \gamma_5 N, \cr
    \mathcal{L}_{S NN}
    &=& - g_{S NN} \bar{N} S N,
    \label{eq:LagMNN}
\end{eqnarray}
where $f_1$, $P$, and $S$ stand $f_1(1285)$, a pseudosclar meson ($\pi(135,0^-)$, $\eta(548,0^-)$), and a scalar meson ($a_0(980,0^+)$,  $f_0(980,0^+)$), and their fields, respectively.
The coupling constant $g_{{\rm M} \gamma \phi  }$ can be calculated from the meson decays as follows.
\begin{eqnarray}
    \Gamma_{f_1 \rightarrow \gamma \phi}
    &=&\frac{k_{\gamma}^3}{12 \pi}
      \frac{M_{\phi}^2}{M_{f_1}^2}
      (M_{f_1}^2 + M_{\phi}^2)g_{f_1 \gamma \phi}^2, \cr 
    \Gamma_{\phi \rightarrow \gamma \Phi}
   & =&\frac{\alpha}{3}\frac{q_{\gamma}^3}{M_{\phi}^2}g_{\Phi \gamma \phi}^2,
    \cr 
    \Gamma_{\phi \rightarrow \gamma S}
   & =&\frac{4\alpha}{3}\frac{q_{\gamma}^3}{M_{\phi}^2}g_{S \gamma \phi}^2.
\end{eqnarray}
Here, $\alpha=e^2/(4\pi)$ and $q_{\gamma}=(M_{\phi}^2-M_{\Phi,S}^2)/(2M_{\phi})$.
By using the values given in the Particle Data Group (PDG) for the  branching ratios~\cite{ParticleDataGroup:2018ovx}, the coupling constants become
\begin{equation}
g_{f_1 \gamma \phi}=0.17\, {\rm GeV}^{-2}, \,\,\,\,
g_{\pi \gamma \phi}=-0.14, \,\,\,\,
g_{\eta \gamma \phi}=-0.71,  \,\,\,\,
g_{a_0 \gamma \phi}=-0.77, \,\,\,\,
g_{f_0 \gamma \phi}=-2.44.
\end{equation}

For the nucleon vertices, from $g_{f_1 NN}= 2.5\pm 0.5$ suggested in Ref.~\cite{Birkel:1995ct} and the values obtained using the Nijmegen potentials~\cite{Stoks:1999bz, Rijken:1998yy}, we use following coupling constants.
\begin{equation}
g_{f_1 NN}=3.0,\,\,\,\, 
g_{\pi NN}=3.0,\,\,\,\,
g_{\eta NN}=6.34,\,\,\,\,
g_{a_0 NN}=4.95,\,\,\,\,
g_{f_0  NN}=-0.51.
\end{equation}
Form factors for both the mesonic and the nucleon vertices are given in the same form as follows.
\begin{equation}
    F_{\rm M}(t)=
    \frac{\Lambda_{\rm M}^2 - M_{\rm M}^2}{
         \Lambda_{\rm M}^2 - t}
\end{equation}

To include not only the exchange of $f_1(1285)$ but also exchanges of higher-spin mesons, we consider $f_1$-Regge which is the Reggeized axial-vector~\cite{Donnachie:2002en,Kochelev:1999zf}. 
Then, the invariant amplitude of $f_1$ from the effective Lagrangians in Eq.~\eqref{eq:LagMVV} and \eqref{eq:LagMNN} needs to change the form. 
Specifically, the Feynman propagator must be substituted with the Regge propagator~\cite{Donnachie:2002en} as follows:
\begin{equation}
    P_{f_1}^{\rm Feyn}(t)=\frac{1}{t-M_{f_1}^2}\,\,\,\,
    \rightarrow\,\,\,\,P_{f_1}^{\rm Regge}(t)=
    \left(\frac{s}{s_{f1}}\right)^{\alpha_{f_1}(t)-1}
    \frac{\pi \alpha_{f_1}'}{{\rm sin}[\pi\alpha_{f_1}(t)]}
    \frac{1}{\Gamma[\alpha_{f_1}(t)]} D_{f_1}(t),
\end{equation}
where the energy-scale factor $s_{f_1}= 1\,{\rm GeV}^2$ and the odd signature factor $D_{f_1}$ is given as
\begin{equation}
D_{f_1}(t) = \frac{{\rm exp}\big[ - i \pi \alpha_{f_1}(t) \big]-1 }{2}.
\end{equation}
The $f_1$-Regge trajectory is $\alpha_{f_1}(t)=0.99 + \alpha_{f_1}'t$ and the slope is chosen as $\alpha_{f_1}'=0.028 \, {\rm GeV^{-2} }$~\cite{Kochelev:1999zf}.

\subsection{$P_s$ contribution}
Here, we define the effective Lagrangians for the $P_s$ for the Feynman diagram Fig.~\ref{fig:01}(c). 
Employing the Rarita-Schwinger formalism for the $J^P=\frac{3}{2}^-$ baryon, we can write the following~\cite{Wang:2019krd,Park:2022nza}.
\begin{eqnarray}
\mathcal{L}_{\gamma N P_s}^{3/2-}&=& e\left[
        \frac{ih_1}{2m_N} \bar{N}\gamma^{\nu} 
        -\frac{h_2}{4m^2_N } \partial^{\nu}\bar{N}
        \right] F_{\mu\nu} P_s^{\mu} + \mathrm{h.c}, \cr
    \mathcal{L}_{P_s \phi N}^{3/2-}
    &=& \, \frac{~i g_{P_s \phi N}^{3/2-}}{2m_N}
    \bar{N}\gamma^{\nu}\phi_{\mu\nu}P_s^{\mu}
    -\frac{g_2}{4m^2_N}
    \partial^{\nu}\bar{N}
    \phi_{\mu\nu} P_s^{\mu}
    +\frac{g_2}{4m^2_N}
    \bar{N}\partial^{\nu}\phi_{\mu\nu}P_s^{\mu} + \mathrm{h.c},
    \label{eq:LagPs}
\end{eqnarray}
where $F_{\mu\nu} = \partial_{\mu}A_{\nu} - \partial_{\nu}A_{\mu}$, and $N$, $A$, $P_s$, and $\phi$ represent the nucleon, photon, $P_s$, and $\phi$ meson fields, respectively. In the present work, we consider only the leading terms of $\mathcal{L}_{\gamma N P_s}^{3/2-}$ and $\mathcal{L}_{P_s \phi N}^{3/2-}$ to avoid additional ambiguities. We adopt $g_{P_s \phi N} = 0.14 + i,0.2$ from Ref.~\cite{Khemchandani:2011et}. For the mass and total width, to reproduce the currently existing data, we use slightly larger mass values, $M_{P_s} = 2.21$ GeV and $\Gamma_{P_s} = 28$ MeV, compared to the original values of 2.071 GeV and 14 MeV, respectively. The electromagnetic (EM) coupling constant $eh_1$ in $\mathcal{L}_{\gamma N P_s}$ can be obtained from the strong coupling constant $g_{P_s \phi N}$ by using the vector meson dominance (VMD) approach similarly to Ref.~\cite{Wang:2019krd,Park:2022nza}. 
\begin{eqnarray}
    eh_{1}&=& g_{P_s \phi N}^{3/2-}\frac{e}{f_{\phi}}
    \frac{2m_N(m_N + m_{P_s})}{(m_{P_s}^2 - m_N^2)m_{\phi}}
    \sqrt{\frac{6m_{\phi}^2 m_{P_s}^2 + m_N^4 - 2 m_N^2 m_{P_s}^2 + m_{P_s}^4}{3m_{P_s}^2 + m_{N}^2}}.
\end{eqnarray}
Here, for the decay constant of the $\phi$ meson $f_{\phi}$, in VMD approach, photons can interact with $P_s$ through $\phi$ vector mesons. 
The interaction Lagrangian for the $\phi$ and the photon can be written as
\begin{equation}
    \mathcal{L}_{\phi \gamma} = - \frac{e m_{\phi}^2}{f_{\phi}} \phi_{\mu}A^{\mu},
\end{equation}
where $f_{\phi}$ is the decay constant of the $\phi$ meson. 
Then, one can determine $e/f_{\phi}$ from the $\phi \rightarrow e^+ e^-$ decay.
\begin{equation}
    \Gamma_{\phi \rightarrow e^+ e^-}
    =\left( 
        \frac{e}{f_{\phi}}
    \right)^2
    \frac{8\alpha |\vec{p}_{e}^{\,cm}|^3}{3m_{\phi}^2},
\end{equation}
where $\vec{p}_{e}^{\,cm}$ denotes the momentum of an electron in the rest frame of the $\phi$ meson.
The $\alpha = e^2/4 \pi$ is the fine-structure constant. 
Using the partial decay width $\Gamma_{\phi \rightarrow e^+ e^-} = 4.249 \, \mathrm{MeV} \times 2.979 \times 10^{-4} = 1.265 \times 10^{-3}\, \mathrm{MeV}$,
\begin{equation}
    \left(\frac{e}{f_{\phi}}\right)^2
    =\frac{3m_{\phi}^2   \Gamma_{\phi \rightarrow e^+ e^-}}{\alpha( m_{\phi}^2 - 4 m_e^2 )^{3/2} } 
    \simeq 0.5101 \times 10^{-4},\,\,\,\,
     \frac{e}{f_{\phi}} \simeq 0.02259.
\end{equation}
Thus, $eh_1$ can be determined as $eh_{1}= 0.00858 + i\,0.01226$.

\subsection{Other nucleon resonances}
Though, there are various resonance states beyond the $\phi N$ threshold $W\sim 1.96$ GeV and their helicity amplitudes of $N^* \rightarrow N\gamma$ decay are well known, the information of $N^* \rightarrow \phi N$ process is very limited. Therefore, to describe the present reaction process, we select the two nucleon-resonance states $N^*(2000,5/2^+)$ and $N^*(2300,1/2^+)$ based on recent CLAS experimental data~\cite{CLAS:2013jlg} and prior theoretical studies~\cite{Kim:2019kef,Kim:2021adl}. The effective Lagrangians of $J^P=\frac{1}{2}^{\pm}$ and $\frac{5}{2}^{\pm}$ for $\gamma NN^*$ vertex for Fig.~\ref{fig:01}(c) can be written as
\begin{eqnarray}
    \mathcal{L}_{\gamma NN^*}^{1/2\pm}
    &=& \frac{e h}{2M_N} \bar{N} \Gamma^{(\mp)}
       \sigma_{\mu\nu}\partial^{\nu} A^{\mu} N^{*} +\mathrm{h.c}, \cr
    \mathcal{L}_{\gamma NN^*}^{5/2\pm}
    &=& e \left[
        \frac{h_1}{(2M_N)^2}\bar{N}\Gamma^{(\mp)}_{\nu}
        -\frac{i h_2}{(2M_N)^3}\partial_{\nu}\bar{N}\Gamma^{(\mp)}
    \right]\partial^{\alpha}F^{\mu\nu}N^{*}_{\mu\alpha} + \mathrm{h.c},
    \label{eq:LagRes1}
\end{eqnarray}
and for $\phi N N^*$ vertex, 
\begin{eqnarray}
   \mathcal{L}_{\phi NN^*}^{1/2\pm}
   & =& \pm \frac{1}{2M_N} \bar{N}
    \left[
        \frac{g_1 M_{\phi}^2}{M_{N^*} \mp M_N }\Gamma^{(\mp)}_{\mu}
        \mp g_2 \Gamma^{(\mp)}_{\mu} \sigma_{\mu\nu}\partial^{\nu}
    \right] \phi^{\mu}N^{*} + \mathrm{h.c},\cr
    \mathcal{L}_{\phi NN^*}^{5/2\pm}
    &=& \left[
        \frac{g_1}{(2M_N)^2} \bar{N}\Gamma^{(\mp)}_{\nu}
        -\frac{ig_2}{(2M_N)^3}\partial_{\nu}\bar{N}\Gamma^{(\mp)}
        +\frac{ig_3}{(2M_N)^3}\bar{N}\Gamma^{(-\mp)}\partial
    \right]\partial^{\alpha}\phi^{\mu\nu} N^{*}_{\mu\alpha} + \mathrm{h.c},
    \label{eq:LagRes2}
\end{eqnarray}
where the notations for $\Gamma^{(\pm)}$ and $\Gamma^{(\pm)}_{\mu}$ are defined as
\begin{equation}
    \Gamma^{(\pm)}
    =\left(\begin{array}{c}
        \gamma_5 \\
        I_{4\times 4}
        \end{array}\right),\quad
    \Gamma^{(\pm)}_{\mu}
    =\left(\begin{array}{c}
        \gamma_{\mu}\gamma_5 \\
        \gamma_{\mu}
        \end{array}\right).
\end{equation}
Here, we consider only $g_1$ to avoid ambiguities due to the lack of information on $N^* \rightarrow \phi N$, and the values of other $g_{2,3}$ are set to be zero for simplicity. From the values of the Breit-Wigner helicities given in PDG data~\cite{ParticleDataGroup:2018ovx} for $N^*(2000,\frac{5}{2}^+)$, the coupling constants of the electromagnetic transition become $h_1^{N^*(2000)}=-4.24$ and $h_2^{N^*(2000)}=4.00$. For other couplings and the decay widths, we follow the values  given in Ref.~\cite{Kim:2019kef}.
\begin{equation}
    h^{N^*(2300)}
    =1.0,\quad g^{N^*(2000)}=4.0, \quad g^{N^*(2300)}=0.1,\,\,\,\,
    \Gamma_{N^*(2000)}=200 \, {\rm MeV},\,\,\,\,
    \Gamma_{N^*(2300)}= 300 \, {\rm MeV}.
\end{equation}
To avoid unreasonably increasing cross sections with respect to $W$ for the nucleon resonance contributions and consider the finite spatial distributions of the hadrons, we use the following Gaussian form factor for $P_s$, $N^*(2000,5/2^+)$, and $N^*(2300,1/2^+)$~\cite{Kim:2017nxg,Kim:2018qfu,Suh:2018yiu,Corthals:2005ce,DeCruz:2012bv}.
\begin{equation}
    F_{N^*}(x)= {\rm exp}\left[-\frac{(x-M_{N^*}^2)^2}{\Lambda_{N^*}^4} \right],\quad x=(s,u).
\end{equation}

\subsection{Direct $\phi$-meson radiation}
The effective Lagrangians for the direct $\phi$-meson radiation contributions, depicted by Fig.~\ref{fig:01}(d), can be written as
\begin{eqnarray}
    \mathcal{L}_{\gamma NN}
    &=& -e\bar{N}\left[
        \gamma_\mu 
        - \frac{\kappa_{N} }{2 M_N}\sigma_{\mu\nu}\partial^{\nu}
    \right]N A^{\mu}, \cr
    \mathcal{L}_{\phi NN}
    &=& - g_{\phi NN} \bar{N}
    \left[
        \gamma_{\mu} 
        - \frac{\kappa_{\phi NN}}{2 M_N}\sigma_{\mu\nu}\partial^{\nu}
    \right]N\phi^{\mu},
    \label{eq:LagDirect}
\end{eqnarray}
where $\kappa_{N} = 1.79$.  
The $\phi NN$ coupling constants are chosen to be $g_{\phi NN} = -0.24$ and $\kappa_{\phi NN} = 0.2$ following the values given in Ref.~\cite{Meissner:1997qt}. 
The individual $s$- and $u$-channel invariant amplitude from Eqs.\eqref{eq:LagDirect} does not satisfy the Ward-Takahashi identity (WTI). 
This problem can be addressed by taking an appropriate sum of those invariant amplitudes with correct form factors~\cite{Ohta:1989ji,Haberzettl:1997jg,Haberzettl:1998aqi,Davidson:2001rk,Kim:2020wrd}. 
For the magnetic parts of the $s$- and $u$-channel amplitudes, which include the second term of $\mathcal{L}_{\gamma NN}$ in Eq.~\eqref{eq:LagDirect}, they satisfy the WTI by themselves. Thus, we can use the following form factor:
\begin{equation}
    F_{N}(x)= 
    \frac{\Lambda_{N}^4}{
         \Lambda_{N}^4 + (x-M_N^2)^2}, \quad x=(s,u).
\end{equation}
For the electric parts, we choose the following common form factor which conserves the on-shell condition and the crossing symmetry~\cite{Davidson:2001rk}.
\begin{equation}
    F_{c}(s,u)= 
    1-[1-F_N(s)][1-F_N(u)].
\end{equation}

\section{Numerical results and Discussions}
In this section, we present the numerical results of the cross sections, angular distributions, SDMEs, and so on,  and discuss them in detail. 
The free parameters in our model are determined as follows. 
Initially, we set $C_{\mathbb{P}} = 6.5$ for the Pomeron, attributing the largest contribution, to account for the cross section behavior at high energies~\cite{Ballam:1972eq,Barber:1981fj}.
This value was also employed in a prior study in which one of the present authors participated~\cite{Kim:2019kef}. 
Following this, we examine the relative phase factors $e^{i \pi \beta}$ for all contributions, with Pomeron's $\beta$ fixed to zero for brevity. 
The relative phases $\beta$ and the cutoffs $\Lambda$ are determined as in Table~\ref{tbl:1}, taking into account the CLAS data for differential cross sections and SDMEs. 
The differential cross section $d\sigma / d{\rm cos} \,\theta$ is obtained using the following equation.
\begin{equation}
    \frac{d\sigma}{d\Omega} 
    = \frac{1}{64 \pi^2 s}\frac{k'}{k}
    \frac{1}{4}\sum_\textrm{spins} 
    \left| 
    \mathcal{M}_{\lambda,\lambda',s,s'}
    \right|^2
\label{eq.diffcs}
\end{equation}
As mentioned in Sec.~\ref{sec:II}, the invariant amplitude $\mathcal{M}$ consists of various t-channel, proton, and nucleon-resonance parts. 
\begin{table}[b]
\begin{tabular}{c||c|c|c|c|c|c|c}
&\hspace{0.4cm} $f_1$\hspace{0.4cm}  & \hspace{0.4cm} PS\hspace{0.4cm}     & \hspace{0.4cm} S  \hspace{0.4cm}   & \hspace{0.4cm} $N$ \hspace{0.4cm} & $N^*(2000)$ & $N^*(2300)$ &\hspace{0.4cm}  $P_s$\hspace{0.4cm}  \\ \hline
Phase $\beta$          & 1     & 0     & 3/2  & 1   & 1   & 1/2 & 1 \\
Cutoff $\Lambda$ [GeV] & 1.5   & 0.87  & 1.35 & 1.0 & 1.0 & 1.0 & 1.0 \\ 
\end{tabular}
\caption{The relative phases $\beta$ and the cutoffs $\Lambda$ for the numerical calculations.}
\label{tbl:1}
\end{table}

\begin{figure}[t]
\includegraphics[width=15cm]{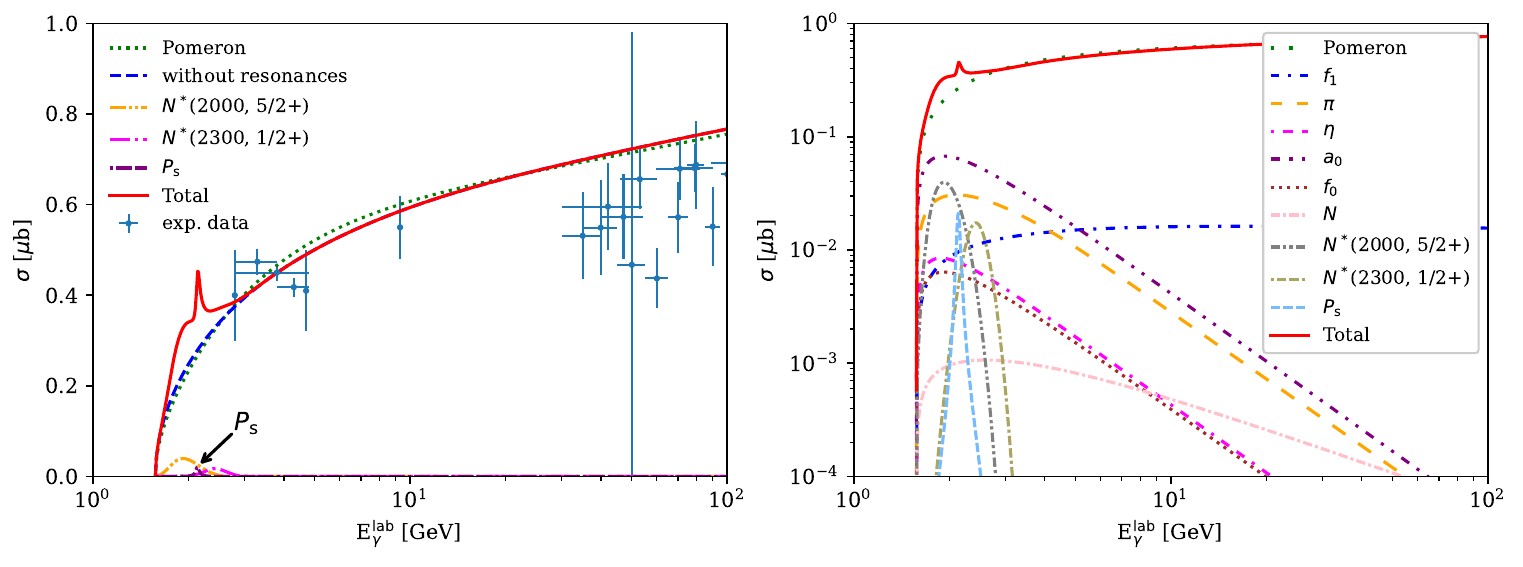}
\caption{The total cross sections for the total and each contribution are plotted in the left panel. The experimental data are taken from Refs.~\cite{Ballam:1972eq,Barber:1981fj,Egloff:1979mg,Busenitz:1989gq}. The total cross section for all contributions is plotted in the logarithmic scale in the right panel for clarity.}
\label{fig:2}
\end{figure}

The numerical results for the total cross section, which is the integral of Eq.~\ref{eq.diffcs} are shown in Fig.~\ref{fig:2}. 
The left panel shows the results of the Pomeron (green dotted), without resonances (blue dashed), $N^*(2000,5/2^+)$ (orange dash-dot-dot), $N^*(2300,1/2^+)$ (pink dash-dot), $P_s$ (violet dot-dash-dash), and the total (red solid) contributions, along with the experimental data~\cite{Ballam:1972eq,Barber:1981fj,Egloff:1979mg,Busenitz:1989gq}. 
The total result of the present work well explains the experimental data up to $10$ GeV~\cite{Ballam:1972eq,Barber:1981fj}, but shows slightly overestimated values than the experimental data in the high-energy region~\cite{Egloff:1979mg,Busenitz:1989gq}. 
It can also be seen that most of the experimental data for the cross section can be explained by the Pomeron contribution alone. 
The right panel shows all contributions in the entire region on the logarithmic scale. 
One can see a bump and peak at E$_{\gamma}^\textrm{ lab} \sim 2$ GeV caused by the nucleon resonances and $P_s$. 
In the right panel, the cross sections of $\pi$ and $a_0$ are comparable to those of the resonances; however, this effect is not very prominent in the total result due to their destructive interference with each other. 
The direct $\phi$-radiation via the proton has almost no contribution across the entire region. 
It can also be seen that most of the contributions other than Pomeron disappear in the high-energy region, but the contribution of $f_1$-Regge remains.

\begin{figure}[t]
\includegraphics[width=15cm]{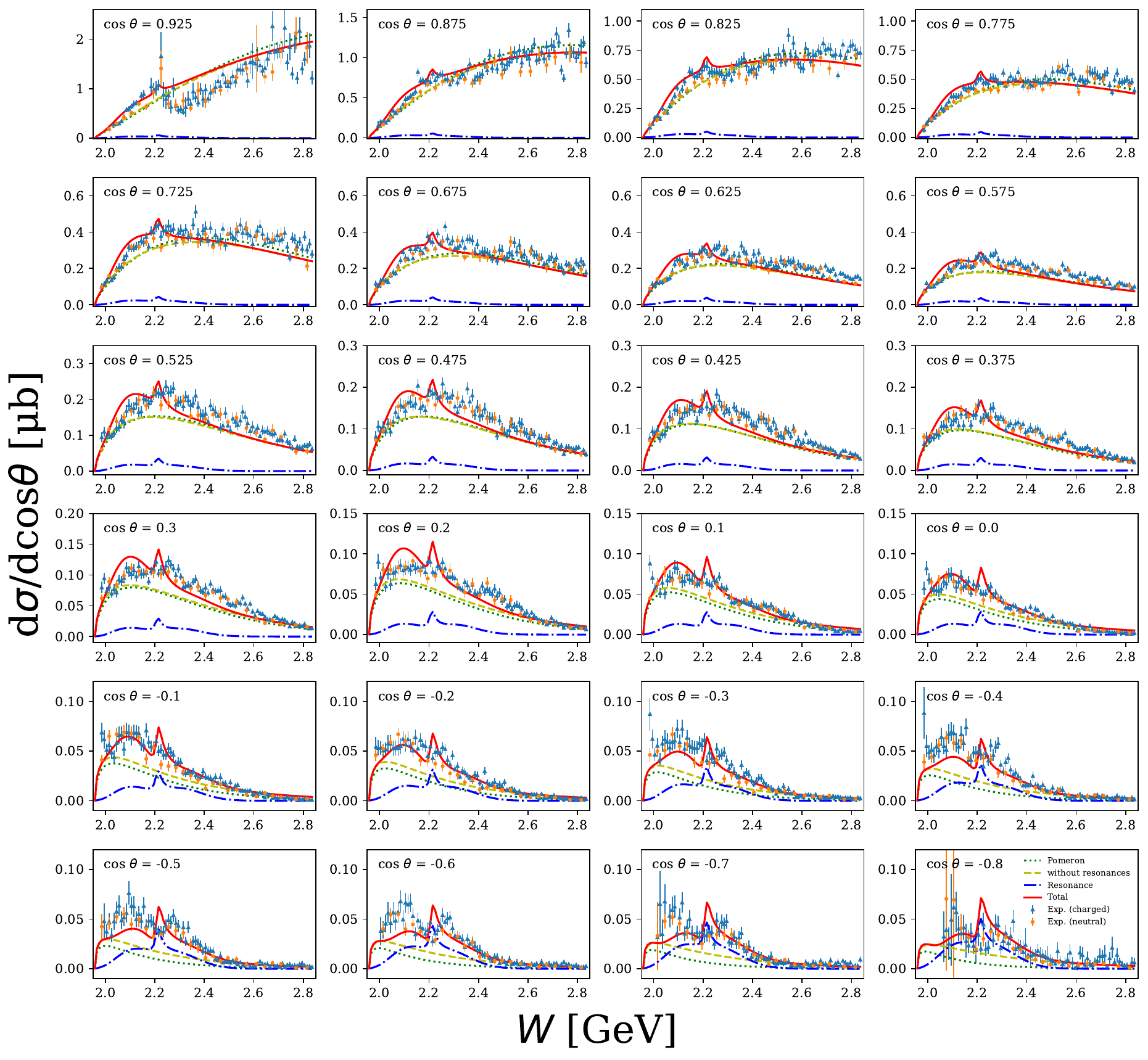}
\caption{The differential cross sections $d \sigma / d {\rm cos} \, \theta$ as a function of $W$. The green dotted line represents the Pomeron contribution, the yellow dashed line represents the resonance-free contribution, the blue dash-dotted line represents the resonance contribution, the red solid line represents the total contribution, and the error bars on triangles and orange circles represent the CLAS experiment data for the charged and neutral $K\bar{K}$ decay modes of $\phi$, respectively.}
\label{fig:3}
\end{figure}
In Fig.~\ref{fig:3}, the results of differential cross section $d\sigma /d\,{\mathrm cos}\,\theta$ as a function of $W$ measured by CLAS experiment~\cite{Dey:2014tfa} are compared with the theory in different ${\mathrm cos}\,\theta$ values. 
The data are obtained for the charged ($\phi \rightarrow K^+ K^-$) and neutral ($\phi \rightarrow K^0_S K^0_L$) decay modes. 
They are measured at uniform angular intervals of 0.05 for $0.375 \le {\rm cos}\, \theta \le 0.925$, and 0.1 for $-0.8 \le {\rm cos}\,\theta \le 0.3$.  
As recommended in Ref.~\cite{Dey:2014tfa}, we compare our numerical result to the experimental data by treating the data of the charged and neutral modes as a unified dataset rather than independent measurements. 
In Fig.~\ref{fig:3}, it can be seen that the contribution of resonances, including $P_s$, plays a crucial role at all angles in reproducing the peak-like structures in the data. 
In particular, at the most forward angle, ${\rm cos}\,\theta = 0.925$, our result is somewhat overestimated for $W > 2.2$ GeV.  
Nevertheless, it partially accounts for the bump observed near $W=2.15$ GeV, attributed to the contribution from resonances, especially $P_s$. 
On the other hand, the contribution of the direct $\phi$-radiation via proton is almost negligible at all angles as mentioned previously. 
At angles of ${\rm cos}\,\theta \geq 0$, the contribution of the Pomeron is dominant while other t-channel contributions are very small. 
However, explaining the experimental data with the Pomeron alone at the angles of ${\rm cos}\, \theta < 0$ is difficult because its contribution becomes smaller and smaller. 
Therefore, the contributions of other $t$ channels and resonances are essential. 

\begin{figure}[t]
\includegraphics[width=15cm]{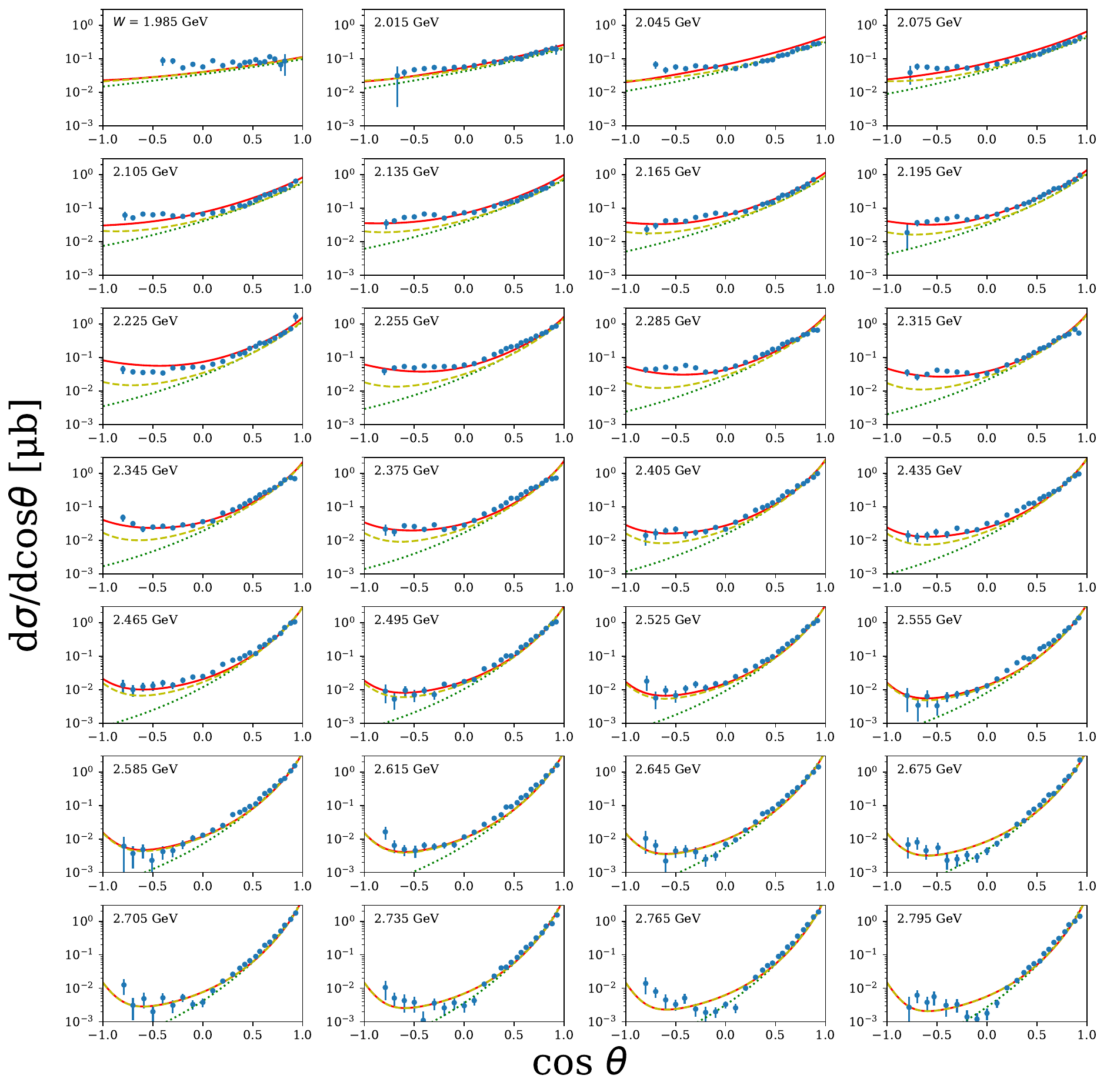}
\caption{The differential cross sections $d \sigma / d {\rm cos} \, \theta$ as a function of ${\rm cos}\,\theta$. 
The notation of the lines and error bars are the same as Fig.~\ref{fig:3}.}
\label{fig:4}
\end{figure}
Fig.~\ref{fig:4} shows the numerical calculation results for $d\sigma/d {\rm cos}\,\theta$ as a function of ${\rm cos}\,\theta$, compared to the CLAS charged-decay experimental data.  
As shown in the results for $W\approx2.5$ GeV or higher, the Pomeron contribution explains the results near the forward region (${\rm cos}\,\theta \sim 1$) well, but the difference between the experimental results and the Pomeron contribution increases as ${\rm cos}\, \theta$ decreases. 
This difference is compensated by other contributions other than the Pomeron and nucleon resonances.
It can be seen that the resonances do not contribute to this region because the result without resonances shows no difference from the result of the total contribution.
However, in the lower $W$ region, the contribution of resonances is very important. In particular, in the vicinity of $W\approx2.2$ GeV, it is difficult to explain the experimental data without resonance contributions, being crucial in the vicinity of cos $\theta < 0$.

\begin{figure}[t]
    \begin{center}
    \includegraphics[width=15cm]{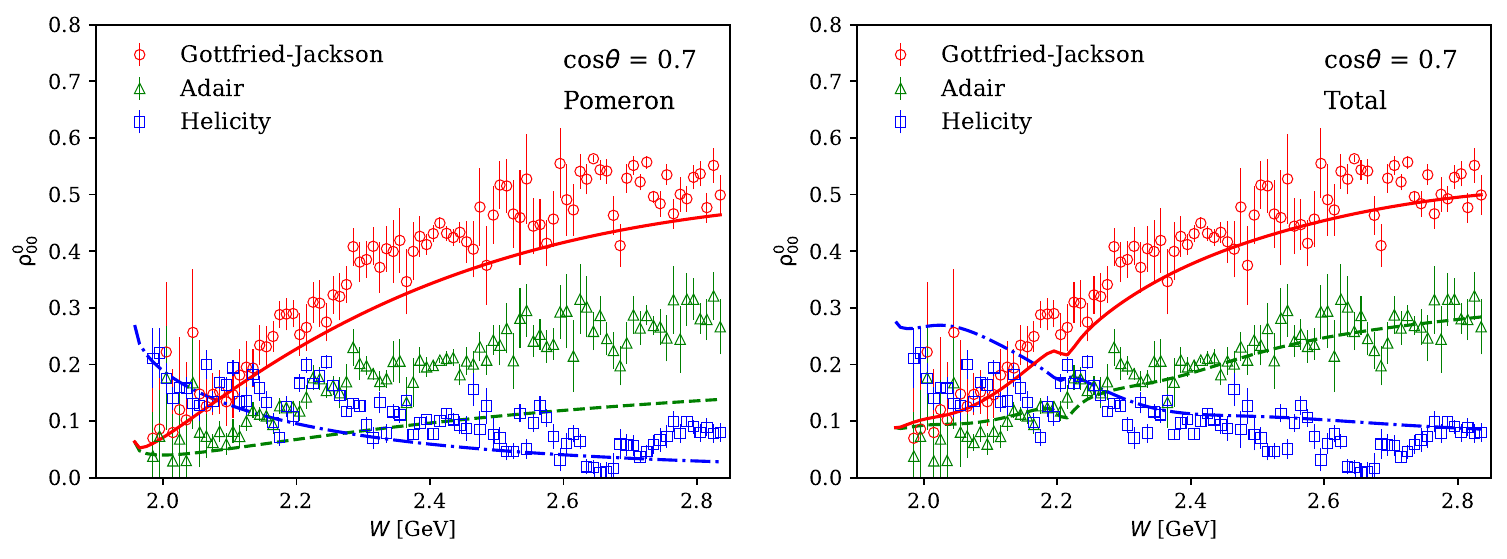}
    \end{center}
    \caption{Spin-density matrix elements (SDMEs) for the Pomeron only (left) and total (right) in the Gottfried Jackson (red solid lines), Adair (blue dot-dashed lines), and Helicity (green dashed) frames.}
\label{fig:5} 
\end{figure}
Fig.~\ref{fig:5} shows the numerical result for the SDME $\rho^0_{00}$ for the Pomeron only (left) and total (right) as a function of $W$ at the angle of ${\rm cos}\, \theta = 0.7$.  
The results are plotted in the Gottfried-Jackson (red solid lines), Adair (blue dot-dashed lines), and Helicity (green dashed) frames, and compared to the CLAS data~\cite{Dey:2014tfa}. 
The definition of $\rho^0_{00}$ is
\begin{equation}
\rho^0_{00} \propto |\mathcal{M}_{\lambda=1, \lambda'=0}|^2 + |\mathcal{M}_{\lambda=-1, \lambda'=0}|^2
\end{equation}
where $\lambda$ and $\lambda'$ are the polarizations of the incoming photon and outgoing $\phi$-meson, respectively. 
For more details about SDMEs and the reference frames, one can refer to Appendix and Ref.~\cite{Schilling:1969um}. 
From this definition, the nonzero value of $\rho^0_{00}$ implies that helicity flip occurs between the photon and the $\phi$-meson. 
Because of this, $\rho^0_{00}$ was often considered when discussing s-channel helicity conservation (SCHC) and $t$-channel helicity conservation (TCHC) along with the properties of the Pomeron~\cite{Kim:2019kef}. 
In the left panel of Fig.~\ref{fig:5}, the numerical results of the Pomeron for $\rho^0_{00}$ are underestimated compared to the experimental data in all three frames.  
This shows that other contributions besides Pomeron are required for the $\phi$-photoproduction.  
In addition, the experimental data shows bump structures with increasing $W$ in each coordinate system.  
By analyzing the difference between the Pomeron results and these bump structures, we can find additional contributions to be considered. 
In the right panel of Fig.~\ref{fig:5} the results for the total contribution are improved and describe the experimental data better. 
In particular, the bump structures at $W\sim1.5$ GeV in each frame are reproduced to some extent and this is due to the contribution of $P_s$. 
It is worth mentioning that the former research without the $P_s$ cannot explain this bump structure. 
Thus, the present SDME data can be experimental evidence for the pentaquark-like state $P_s$.

As mentioned above, to explain the behavior in the cross sections, including the bump structure in forward angles, and the SDMEs in the present work, contributions from the processes of $N^*(2000,\frac{5}{2}^+)$, $N^*(2300,\frac{1}{2}^+)$, and $P_s(2210,\frac{3}{2}^-)$ are essential.  However, in reality, additional resonance states not currently considered may contribute. Moreover, the fitted parameters such as the relative phases, couplings, widths, masses, and the spin-parity of the nucleon resonances can vary, because of interferences with the additional resonances. Therefore, further investigation is required to eliminate such uncertainties together with more reliable experimental data. For example, an experiment using the EIC
to be built at the Brookhaven National Laboratory (BNL)~\cite{AbdulKhalek:2021gbh} will be the best to investigate the electromagnetic production of $\phi$ meson, exploring the existence of $P_s$.

Theoretical research on a similar reaction, particularly electroproduction of the hidden-flavor vector meson $J/\Psi$, has already been conducted in Ref.~\cite{Park:2022nza}.  
In Ref.~\cite{Park:2022nza}, the differential cross section for the production of $J/\psi$ meson via $P_c$ was evaluated as a function of the pseudorapidity and the transverse momentum in the collider kinematics.  
Using the same formalism, the total cross section for $P_s$ production turns out to be 4.09 pb in $e$+$p$ collisions at the EIC's peak energy ($W = 126$ GeV). 
Given the luminosity of 10 fb$^{-1}$ per month, this translates to the potential measurement of about 40k $P_s$ particles annually. 
In this photo-production process, $P_s$ has similar rapidity with the proton beam, necessitating a dedicated detector positioned at the very forward (proton-going) direction to effectively measure these. 
Furthermore, as observed in Ref.~\cite{Park:2022nza}, considering polarized cross sections, the spin-parity of $P_c$ can be distinguished. 
Similarly, we can theoretically predict polarized cross sections for the nucleon resonances including $P_s$, and comparing them with future experimental results, is anticipated to enhance our understanding of the $\phi$-photoproduction with the exotic baryon $P_s$.

\section{Summary}
We explore the photoproduction reaction of $\phi$-mesons utilizing the effective Lagrangian method within the tree-level Born approximation. 
Our investigation encompasses contributions from various contributions, including the Pomeron, $f_1$-Regge, pseudoscalar particles ($\pi$, $\eta$), scalar particles ($a_0$, $f_0$), protons, and three-nucleon resonance states. 
In addition to the $N^{*}(2000,5/2^+)$ and $N^{*}(2300,1/2^+)$ states considered in prior theoretical studies~\cite{Kim:2019kef}, we introduce the contribution of the pentaquark-like nucleon-resonance state $P_s$, anticipated to be a $K^* \Sigma$-bound state, as suggested in Ref.~\cite{Khemchandani:2011et}. 
To refine our model, we determine unknown parameters, including the Pomeron parameter $C_{\mathbb{P}}$, relative phases between contributions denoted as $\beta$, and cutoffs $\Lambda$. 
This determination is achieved through a comparison with early total cross section data~\cite{Ballam:1972eq,Barber:1981fj,Egloff:1979mg,Busenitz:1989gq} and recent data from the CLAS experiment~\cite{Dey:2014tfa}.

The numerical findings from this investigation reveal that, in terms of the total cross section, contributions other than the Pomeron are generally minor, particularly except for behavior near the threshold. 
At higher energies, contributions other than the Pomeron and $f_1$ Regge are deemed negligible. 
When focusing on the production reaction at specific angles rather than the total cross section, relying solely on the Pomeron contribution becomes inadequate. 
In such cases, analyzing additional contributions becomes necessary. 
Consequently, we compute $d\sigma /d {\rm cos}\,\theta$ to assess various angles and center-of-mass energies, providing a comparison with CLAS data. 
From this analysis, it becomes evident that the Pomeron contribution alone effectively accounts for the experimental data in the vicinity of ${\rm cos} \,\theta \sim 1$. 
However, as ${\rm cos}\,\theta$ decreases, the deviation from the data increases, highlighting the necessity of contributions other than the Pomeron, particularly at lower ${\rm cos}\,\theta$ values. 
Crucially, the inclusion of $P_s$ and other nucleon resonances proves essential for accurately describing the observed bump near $W \sim 2.15$ GeV in very forward angles, as well as capturing the behaviors within the range of $(2.0 -2.3)$ GeV. 
Additionally, in the region for $W \ge 2.5$ GeV, where nucleon resonances are nearly absent, contributions from t-channel mesons, distinct from the Pomeron, become necessary.

In conclusion, we compute $\rho^0_{00}$ at ${\rm cos}\,\theta = 0.7$ in the Gottfried-Jackson, Adair, and Helicity frames, subsequently comparing these calculations with CLAS data. 
When solely accounting for the Pomeron, the results consistently fall below the data in all frames. 
However, with the inclusion of all contributions, this discrepancy is notably diminished, and the data aligns well with the model. 
Notably, a significant accomplishment of this study is the successful reproduction of the observed bump at $W \sim 2.15$  GeV through the incorporation of $P_s$. 
Our study demonstrates that the upcoming EIC experiment will enable high-precision measurements of the $P_s$ resonance with the luminosity of 10 fb$^{-1}$ per month.
More detailed works on $P_s$ production in EIC are in progress and will appear elsewhere.

\section*{Acknowledgment}

The authors are grateful for the fruitful discussions with Sang-Ho Kim (Soongsil University), Kanchan Pradeepkumar Khemchandani (Federal University of São Paulo), and Alberto Martinez Torres (University of São Paulo). 
This work is supported by the National Research Foundation of Korea (NRF) grants funded by the Korean government (MSIT) (2018R1A5A1025563). 
The work of Y.S.K. was supported by the NRF grant (2022R1A2C1011549). 
The work of S.i.N. was also supported by the NRF grants (2022R1A2C1003964 and 2022K2A9A1A06091761).

\section*{Appendix}
\label{sec:AppA}
\begin{figure}[h]
    \begin{center}
    \includegraphics[width=8cm]{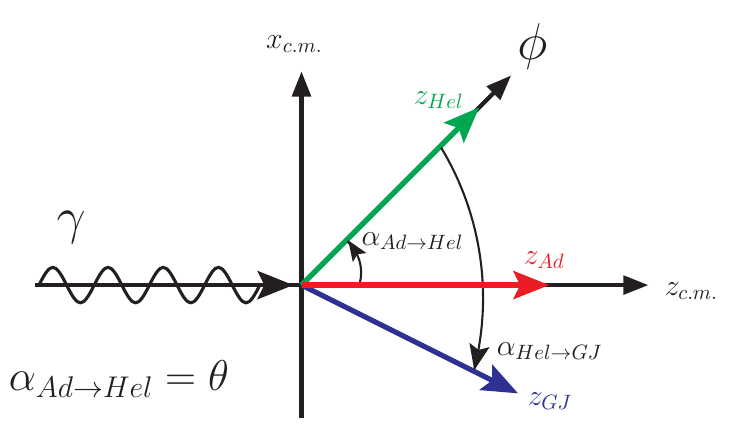}
    \end{center}
    \caption{The spin-quantization axes for the helicity ($z_\textrm{Hel}$, in green), Adair ($z_\textrm{Ad}$, in red), and Gottfried-Jackson ($z_\textrm{GJ}$, in blue) frames, in the c.m. frame.}
\label{fig:6}
\end{figure}

Here we briefly discuss the SDMEs and helicity conservation.  
The spin density matrix element (SDMEs) is defined in terms of the Helicity amplitude as follows~\cite{Schilling:1969um}:
\begin{eqnarray}
\rho^0_{\lambda'_1 \lambda'_2} 
&=& \frac{1}{N}
    \sum_{s,s',\lambda}
    \mathcal{M}_{s,s',\lambda, \lambda'_1} \mathcal{M}^*_{s,s',\lambda, \lambda'_2},
    \cr
\rho^1_{\lambda'_1 \lambda'_2}
&=&\frac{1}{N}
    \sum_{s,s',\lambda}
    \mathcal{M}_{s,s',\lambda, \lambda'_1} \mathcal{M}^*_{s,s',\lambda, \lambda'_2},
    \cr
\rho^2_{\lambda'_1 \lambda'_2}
&=& \frac{1}{N}
    \sum_{s,s',\lambda}
    \mathcal{M}_{s,s',\lambda, \lambda'_1} \mathcal{M}^*_{s,s',\lambda, \lambda'_2},
    \cr
\rho^3_{\lambda'_1 \lambda'_2}&=& \frac{1}{N}
    \sum_{s,s',\lambda}
    \mathcal{M}_{s,s',\lambda, \lambda'_1} \mathcal{M}^*_{s,s',\lambda, \lambda'_2},
\end{eqnarray}
where the normalization factor is $N=\sum|\mathcal{M}_{s,s',\lambda,\lambda'}|^2$.  Because the SDMEs are not Lorentz invariant quantities, the values change depending on the choice of the spin-quantization direction for the $\phi$ meson. There are three reference frames often used, the Adair frame, the helicity frame, and the Gottfried-Jackson frame, as shown in Fig.~\ref{fig:6}.

The spin-quantization axis $z_\textrm{Ad}$ coincides with the z-axis for the c.m. frame. Since $z_\textrm{Hel}$ points along the direction of the $\phi$ meson in the c.m. frame, the angle between the helicity and Adair frames is $\theta$.  In the Gottfried-Jackson frame, $z_\textrm{GJ}$ is equal to the direction of the incoming photon in the $\phi$-meson rest frame. The $z_\textrm{Hel}$ and $z_\textrm{GJ}$ are favored for discussion of the s-channel and t-channel helicity conservations, respectively. The SDMEs in one reference frame can be transformed to them in other frames by a Wigner rotation. The rotation angle $\alpha $'s are expressed as in Ref.~\cite{Schilling:1969um}
\begin{eqnarray}
    \alpha_{\textrm{Ad} \rightarrow \textrm{Hel}}
    &=&\theta, \cr
    \alpha_{\textrm{Hel} \rightarrow \textrm{GJ}}
    &=&- {\rm cos}^{-1} 
           \left(\frac{v - {\rm cos}\, \theta}{v\, {\rm cos}\, \theta -1 } \right), 
           \cr
    \alpha_{\textrm{Ad} \rightarrow \textrm{GJ}}
    &=& \alpha_{\textrm{Ad} \rightarrow \textrm{Hel}}+\alpha_{\textrm{Hel}\rightarrow \textrm{GJ}},
\end{eqnarray}
where $v$ is the velocity of the $\phi$ meson in the c.m. frame.


\end{document}